\documentclass[journal]{IEEEtran}
\usepackage{soul,framed} %,caption
\usepackage[pdftex,dvipsnames]{xcolor} 
\usepackage{color}
\usepackage[pdftex]{graphicx}
\graphicspath{{../pdf/}{../jpeg/}}
\DeclareGraphicsExtensions{.pdf,.jpeg,.png}
\usepackage[cmex10]{amsmath}
\usepackage{array}
\usepackage{mdwmath}
\usepackage{amssymb}
\usepackage{mdwtab}
\usepackage{eqparbox}
\usepackage{url}
\usepackage{longtable}
\usepackage{multirow}
\usepackage{tabularx}
\usepackage{footnote}
\usepackage{threeparttable}
\usepackage{booktabs}
\usepackage{cite}
\usepackage{siunitx}
\usepackage{adjustbox}
\usepackage{xspace}
\makesavenoteenv{tabular}
\makesavenoteenv{table}
\usepackage{xargs} 
\newcommand{\etal}{\emph{et al.}\xspace}

\newcommand{\mat}[1]{\boldsymbol{#1}} % matrix
\usepackage[colorinlistoftodos,textsize=scriptsize,textwidth=8mm, disable]{todonotes}

\usepackage{marginnote}

\newcommandx{\unsure}[2][1=]{\todo[linecolor=red,backgroundcolor=red!25,bordercolor=red,#1]{#2}}
\newcommandx{\change}[2][1=]{\todo[linecolor=blue,backgroundcolor=blue!25,bordercolor=blue,#1]{#2}}
\newcommandx{\info}[2][1=]{\todo[linecolor=OliveGreen,backgroundcolor=OliveGreen!25,bordercolor=OliveGreen,#1]{#2}}
\newcommandx{\improvement}[2][1=]{\todo[linecolor=Plum,backgroundcolor=Plum!25,bordercolor=Plum,#1]{#2}}
\newcommandx{\thiswillnotshow}[2][1=]{\todo[disable,#1]{#2}}



\newcommand{\infol}[2][]{{%
 \let\marginpar\marginnote
 \reversemarginpar\info[#1]{{\bf #2}}}}

\begin{document}
\title{Cine Cardiac MRI Motion Artifact Reduction Using a Recurrent Neural Network}

\author{Qing Lyu,~\IEEEmembership{Student Member,~IEEE}, Hongming Shan,~\IEEEmembership{Member,~IEEE}, Yibin Xie, Debiao Li$^\ast$, and Ge Wang$^\ast$, ~\IEEEmembership{Fellow,~IEEE}
\thanks{Asterisk indicates corresponding authors.}
% \thanks{This work was partially support by NIH/NCI under award numbers R01CA233888 and  R01CA237267, and NIH/NIBIB under award number R01EB026646.}
\thanks{Q. Lyu, H. Shan, and G. Wang are with Biomedical Imaging Center, Department of Biomedical Engineering/School of Engineering/Center for Biotechnology and Interdisciplinary Studies, Rensselaer Polytechnic Institute, Troy, NY 12180, USA (e-mail: lyuq@rpi.edu; hmshan@ieee.org; wangg6@rpi.edu).}
\thanks{Y. Xie is with Biomedical Imaging Research Institute, Cedars-Sinai Medical Center, Los Angeles, CA 90048, USA (e-mail: Yibin.Xie@cshs.org).}
\thanks{D. Li is with Biomedical Imaging Research Institute, Cedars-Sinai Medical Center, Los Angeles, CA 90048, USA, Department of Medicine, University of California, Los Angeles, CA 90095, USA, and Department of Bioengineering, University of California, Los Angeles, Los Angeles, CA 90095, USA (e-mail: Debiao.Li@cshs.org).}
}

\maketitle

\begin{abstract}
Cine cardiac magnetic resonance imaging (MRI) is widely used for diagnosis of cardiac diseases thanks to its ability to present cardiovascular features in excellent contrast. As compared to computed tomography (CT),  MRI, however, requires a long scan time, which inevitably induces motion artifacts and causes patients' discomfort. Thus, there has been a strong clinical motivation to develop techniques to reduce both the scan time and motion artifacts. Given its successful applications in other medical imaging tasks such as MRI super-resolution and CT metal artifact reduction, deep learning is a promising approach for cardiac MRI motion artifact reduction. In this paper, we propose a recurrent neural network to simultaneously extract both spatial and temporal features from under-sampled, motion-blurred cine cardiac images for improved image quality. The experimental results demonstrate substantially improved image quality on two clinical test datasets. Also, our method enables data-driven frame interpolation at an enhanced temporal resolution. Compared with existing methods, our deep learning approach gives a superior performance in terms of structural similarity (SSIM) and peak signal-to-noise ratio (PSNR).
\end{abstract}

\begin{IEEEkeywords}
Cardiac magnetic resonance imaging (MRI), motion artifact reduction, fast MRI, recurrent neural network.
\end{IEEEkeywords}

\IEEEpeerreviewmaketitle

\section{Introduction}
\IEEEPARstart{M}{agnetic} resonance imaging (MRI) is one of the most extensively used medical imaging modalities for diagnosis and intervention. It can be flexibly programmed to noninvasively map anatomical structures in a patient with different types of tissue contrasts without ionizing radiation. Currently, cine cardiac MRI is used to image a beating heart and evaluate cardiac functions and vascular abnormalities~\cite{motwani2013mr}. Generally, the cine cardiac MRI scan is synchronized with the cardiac cycle signal, known as the electrocardiogram (ECG), to suppress motion artifacts~\cite{sechtem1987cine}. For a typical cine cardiac MRI scan, MR data in multiple heart cycles are continuously recorded and retrospectively sorted into different phases to form consistent datasets for each of these phases. The resultant MR imaging process turns out to be complicated and time consuming. 

High-quality cine cardiac MR images are a key component for accurate diagnosis~\cite{leiner2019machine}, which can currently only be obtained using a long scan. However, patient’s motion happens during such a long process; for example, due to respiration and bulk body movement, inevitably introducing motion artifacts out of the ECG synchronization. Also, a long scan time increases patient's discomfort and scan expense. To solve these problems, in recent years researchers proposed various methods to simplify cardiac MRI, speed up scanning, and reduce motion artifacts. Self-gated cine cardiac MRI was proposed to acquire cardiac MR images in real-time without using ECG~\cite{larson2004self, crowe2004automated}. However, it suffers from complicated reconstruction steps and lowers spatial and temporal resolution. Schmitt~\etal built a 128‐channel receive‐only cardiac coil for highly accelerated cardiac MRI~\cite{schmitt2008128}. In post-processing studies, Tsao~\etal proposed \textit{k-t} BLAST (Broad-use Linear Acquisition Speed-up Technique) and \textit{k-t} SENSE (SENSitivity Encoding) methods that utilize spatial-temporal correlation among \textit{k}-space data as a function of time to boost cardiac MRI image quality and improve diagnostic performance~\cite{tsao2003k}.  Lingala~\etal~\cite{lingala2011accelerated}, Tr{\'e}moulh{\'e}ac~\etal~\cite{tremoulheac2014dynamic}, Otazo~\etal~\cite{otazo2015low}, and Christodoulou~\etal~\cite{christodoulou2018magnetic} developed compressed sensing MRI algorithms, based on low-rank and sparsity, to use sparse data for accelerated cardiac MRI data acquisition. Schloegl~\etal~\cite{schloegl2017infimal} and Wang~\etal~\cite{wang2020dynamic} proposed total generalized variation of spatial-temporal regularization for MRI image reconstruction.

\begin{figure*}[!hbt]
\centering
\includegraphics[width=7in]{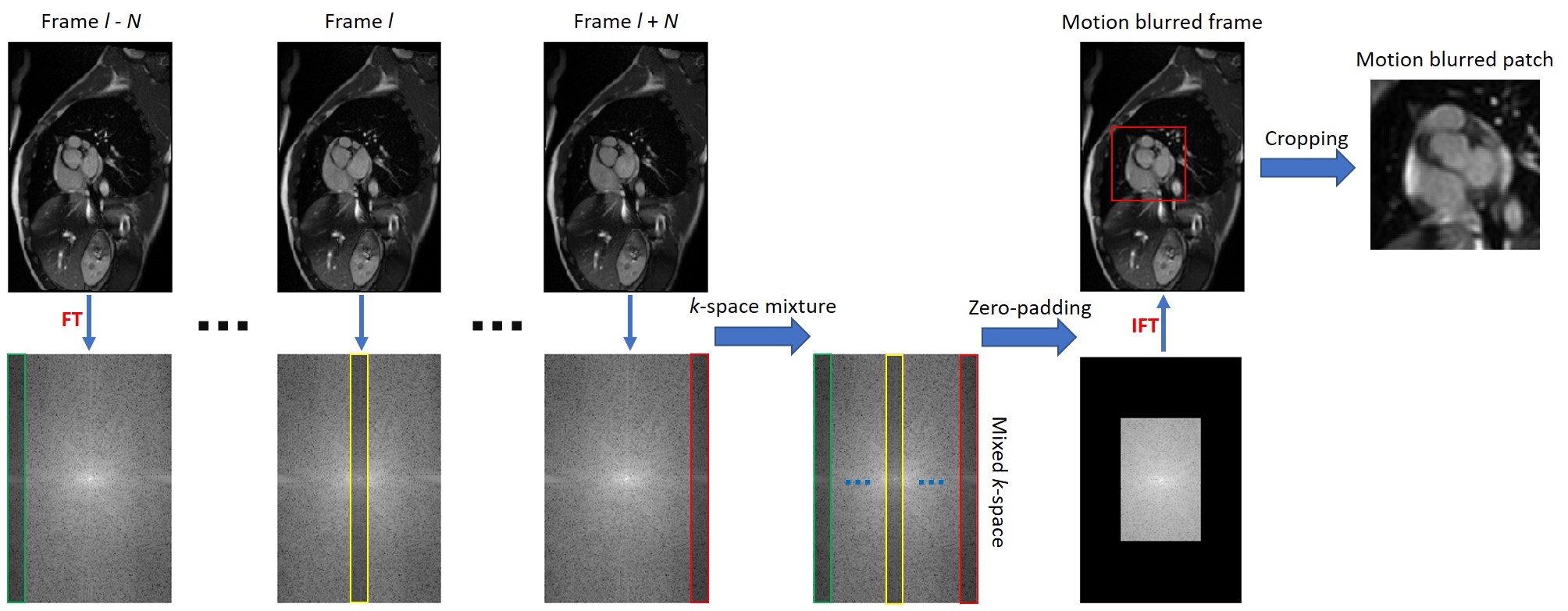}
\caption{Mixture of \textit{k}-space data points from adjacent frames to create motion blurred MRI images. The Fourier transform converts each motion-free frame into the \textit{k}-space. A few frequency encoding data lines are selected from each \textit{k}-space and put together forming a mixed \textit{k}-space. A motion-blurred frame is generated after converting zero-padded mixed \textit{k}-space into the image domain.}
\label{fig_1}
\end{figure*}

In recent years, deep learning has been successfully used in post-processing static MR images for various tasks such as segmentation~\cite{poudel2016recurrent, tan2017convolutional, romaguera2017left, luo2016deep}, registration~\cite{de2017end,shan2020synergizing}, and super-resolution~\cite{lyu2019super,chaudhari2018super, chen2018efficient, lyu2020multi, lyu2020mri, liu2018fusing, zeng2018simultaneous}. In addition 
to applying deep learning to static images, neural networks also have a strong ability in analyzing dynamic information; for example, video compression artifact reduction~\cite{yang2018multi, guan2019mfqe, xue2019video} and dynamic cardiac MRI reconstruction~\cite{schlemper2017deep,qin2018convolutional}. Neural networks have also been used on medical imaging modalities to aid diagnosis. Luo~\etal~\cite{luo2016novel} proposed a neural network for ventricular volume prediction without any segmentation. Zhang~\etal~\cite{zhang2019spatio} adapted a convolutional long short-term memory (ConvLSTM) to predict the growth of tumors. Bello~\etal~\cite{bello2019deep} trained a fully convolutional network based on anatomical priors to analyze cardiac motion and predict patient's survival. Based on all these studies, in this paper we design an advanced deep recurrent neural network and demonstrate the feasibility of using deep learning to reduce motion artifacts and blurring in cine cardiac MR images.

In our study, we mix \textit{k}-space data within several adjacent frames in a cine cardiac MRI sequence and adopt the zero padding under-sampling strategy to simulate the acquisition process under the ECG-free fast cine cardiac MRI scan protocol. Then, we propose a deep recurrent neural network with two ConvLSTM branches to dynamically process cine images and generate motion-compensated results comparable to those from an ECG-gated cine cardiac MRI scan. After training, the proposed network is tested on two independent clinical datasets, demonstrating superior quality and high robustness. Remarkably, given incomplete sequences, our method is able to predict missing frames in high fidelity via frame interpolation, indicating the feasibility of improving temporal resolution of the cine cardiac MRI scan.

\section{Methodology}
The overall goal of this study is to design an advanced neural network to reduce motion artifacts in a cine cardiac MRI image sequence acquired in a fast  scan. To achieve this goal, we first create motion blurred images based on a high-quality cine cardiac MRI dataset, and then train a deep recurrent network in a supervised fashion to learn how to remove motion artifacts and improve image quality. Apart from reducing motion artifacts on each frame in a complete sequence, the proposed network can also predict missing frames in incomplete sequences.

\begin{figure*}[!hbt]
\centering
\includegraphics[width=7in]{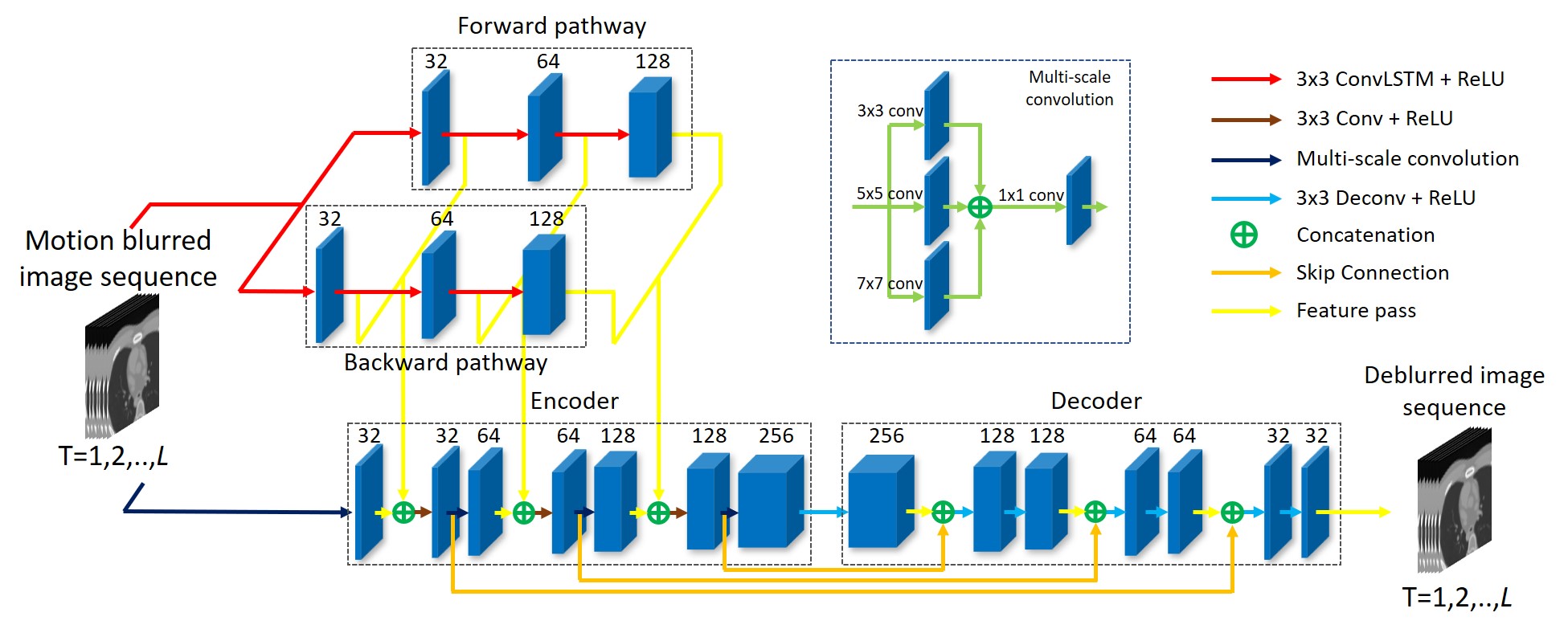}
\caption{Architecture of the proposed deep recurrent neural network. It includes two ConvLSTM branches and an encoder-decoder sub-network. Multi-scale structures are used in the encoder, shown by dark blue arrows.}
\label{fig_2}
\end{figure*}

\subsection{Datasets}
\subsubsection{ACDC dataset}
The well-known automated cardiac diagnosis challenge (ACDC) dataset~\cite{bernard2018deep} is selected in this study. It includes 150 cardiac MRI scans from 5 subgroups: normal, patients with previous myocardial infarction, patients with dilated cardiomyopathy, patients with hypertrophic cardiomyopathy, and patients with abnormal right ventricle. Each subgroup contains 30 scans. The number of slices in one scan ranges from 6 to 21 with 12 to 35 phases. All scans were collected from two MRI scanners (1.5T Siemens Area and 3.0T Siemens Trio Tim). The slice thickness is 5 or 8 mm, and spatial resolution goes from 1.37 to 1.68 $mm^2/pixel$. Among all these scans, we found 69 scans with 30 phases and randomly separated them into two groups: 50 scans with a total of 372 short-axis slices for training and the remaining 19 scans with a total of 140 slices for testing.

\subsubsection{Cedars dataset}
Independent to the ACDC dataset, the Cedars dataset includes 9 Cedars cine cardiac MRI scans acquired from the Cedars-Sinai medical center via Siemens Avanto 1.5T MRI scanner. Each scan has 8 to 13 short-axis slices with 9 to 30 phases. The slice thickness is 8 mm. Spatial resolution is different among each subject, ranging from 1.406 to 1.734 $mm^2/pixel$. Among all 80 slices, we randomly selected 10 slices for fine-tuning and used the remaining 70 slices for testing.

\subsection{Motion blurring}
All original sequences from the ACDC and Cedars datasets were acquired by MRI scanners in conjunction with ECG-gated signals, which are assumed as the motion-free ground truth. Fig.~\ref{fig_1} depicts the creation process of motion blurred sequences. To produce motion-blurred sequences, we first converted each ground truth image in a gated MRI sequence into a \textit{k}-space array using the 2D Fourier transform. Then, for a given sequence with $2N+1$ frames, several frequency encoding data lines were extracted from each \textit{k}-space. These extracted data lines were fused together to form a mixed \textit{k}-space array. By such a design, data points from adjacent frames at different cardiac phases contribute to the formation of a motion-blurred image, which simulates the actual scan process with relevant pulse sequences such as steady-state free precession (SSFP) or echo planar imaging (EPI). To verify our assumption that our method can boost a fast MRI scan, we further adopted the zero-padding under-sampling strategy and only kept central 25\% data. Finally, the motion blurred frames were obtained after the 2D inverse Fourier transform that converts the down-sampled mixed \textit{k}-space data back into the image domain. Through this motion blurring process, we can realistically produce motion blurred cine cardiac MRI sequences that can be seen using an ECG-free cardiac MRI scan protocol. The overall process can be expressed as
\begin{equation}\label{equ_1}
    \widehat{\mat{I}}_{l} = \mathcal{F}^{-1}\left[\phi\left(\sum_{i=l-N}^{l+N} \psi(\mathcal{F}(\mat{I}_{i}))\right)\right],
\end{equation}
where $\widehat{\mat{I}}$ and $\mat{I}$ stand for motion blurred and motion free images respectively, $\mathcal{F}$ and $\mathcal{F}^{-1}$ indicate the 2D Fourier transform and inverse transform respectively, $\psi$ selects \textit{k}-space data lines, $\sum$ mixes the \textit{k}-space data segments from $2N+1$ adjacent frames, and $\phi$ represents zero padding.

In this study, we only focused on the cardiac region that suffers from severe motion artifacts, which is a cropped patch of size 100 $\times$ 100 pixel containing cardiac structures in each pair of motion blurred and motion free images. All pixel values were normalized to the range of [0, 1].

\subsection{Neural network}
We designed a deep recurrent neural network serving as the generator in the  generative adversarial network (GAN) framework with Wasserstein distance and gradient penalty, which includes an encoder-decoder network and two ConvLSTM branches, as shown in Fig.~\ref{fig_2}. 

For the two ConvLSTM branches, there is a forward branch and a backward branch, both of which use identical three-layer ConvLSTM structures. The two branches can both extract forward and backward temporal features. Different from traditional LSTM that only utilizes temporal information, ConvLSTM~\cite{xingjian2015convolutional} combines traditional LSTM with spatial convolution so that spatial and temporal information is taken into synergistic account. In this study, we set the size of convolution kernels to 3 $\times$ 3 so that local spatial features can be extracted. A ConvLSTM layer can be expressed as
\begin{align}\label{equ_2}
    i_{t} & = \sigma(\mat{W}_{xi}\ast \mat{X}_{t} + \mat{W}_{hi}\ast \mat{H}_{t-1} + \mat{W}_{ci}\circ \mat{C}_{t-1} + b_{i}) \notag\\ 
    f_{t} & = \sigma(\mat{W}_{xf}\ast \mat{X}_{t} + \mat{W}_{hf}\ast \mat{H}_{t-1} + \mat{W}_{cf}\circ \mat{C}_{t-1} + b_{f}) \notag\\
    \mat{C}_{t} & = f_{t}\circ \mat{C}_{t-1} + i_{t}\circ \tanh({\mat{W}_{xc}\ast \mat{X}_{t}+\mat{W}_{hc}\ast \mat{H}_{t-1}+b_{c}}) \notag\\ 
    o_{t} & = \sigma(\mat{W}_{xo}\ast \mat{X}_{t} + \mat{W}_{ho}\ast \mat{H}_{t-1} + \mat{W}_{co}\circ \mat{C}_{t} + b_{o}) \notag\\
    \mat{H}_{t} & = o_{t}\circ \tanh({\mat{C}_{t}}),
\end{align}
where $\ast$ denotes a 2D convolution, $\circ$ stands for the Hadamard product, $i_t$, $f_t$, $\mat{C}_t$, $o_t$, and $\mat{H}_t$ are 3D tensors representing input gate, forget gate, cell state, output gate, and hidden state respectively. The stride and padding steps were set to 1. The numbers of hidden state channels in each ConvLSTM layer are 32, 64, and 128.

The proposed network relies on an encoder-decoder network. The encoder contains seven convolution blocks and gradually extracts spatial features from inputs. A multi-scale structure is used in each convolution block with 3 types of convolution kernels, and spatial features can be extracted locally and globally. Features of each frame extracted from the two ConvLSTM branches are added into the encoder and concatenated with corresponding features after the first, third, and fifth convolution blocks. Such a feature combination ensures both spatial and temporal features extracted by the ConvLSTM branches to be fully utilized and encoded by the encoder. All the convolution layers in the encoder use a stride of 1 and a padding of 1. The decoder has seventh decovolution layers and receives features extracted by the second, fourth, sixth, and seventh convolution blocks. The decoder gradually decodes features and finally generates motion deblurred sequences. All the deconvolution layers in the decoder use a kernel size of 3, a stride of 1, and a padding of 1. In the encoder-decoder network, there is a ReLU activation function right after each convolution or deconvolution layer.

The discriminator contains six convolution layers and two fully connected layers. All convolution layers use 3 $\times$ 3 convolution kernels with a stride of 1 and padding of 1. The numbers of convolution kernels in these layers are 64, 64, 128, 128, 256, and 256 respectively. There are ReLU activation functions after convolution, and then a max-pooling layer with a size of 2 $\times$ 2 and a stride of 2 only after the second and fourth convolution layers respectively. There are 1,024 neurons in the first fully convolutional layer but only 1 neuron in the final layer.

\subsection{Loss function}
We use the perceptual loss~\cite{johnson2016perceptual} as part of the objective function for the generator, which is based on the pre-trained VGG16 model~\cite{simonyan2014very}. The $\ell_2$ norm of features from the second, fourth, seventh, and tenth convolutional layers are used for loss calculation. The objective function is the summation of all perceptual losses over all frames in a sequence:
\begin{equation}\label{equ_3}
    \mathcal{L}_{\rm per} = \sum_{l=1}^{L}\sum_{i=2, 4, 7, 10} \mathbb{E}\Big\|\theta_{i}(\widehat{\mat{I}}_{l})-\theta_{i}(\mat{I}_{l})\Big\|_{2}^{2},
\end{equation}
where $\theta_{i}$ is the feature extracted by the $i$-th convolutional layer in the pre-trained VGG16 network, and $L$ stands for the length of the input sequence. The overall objective function for the generator is the combination of the perceptual loss and the adversarial loss:
\begin{equation}\label{equ_4}
    \mathcal{L}_{G} = -\sum_{l=1}^{L}\mathbb{E}\Big(D(G(\widehat{\mat{I}_{l}}))\Big) + \lambda_{\rm per}\mathcal{L}_{\rm per},
\end{equation}
where $\lambda_{\rm per}$ is empirically set to 0.1.

There are two benefits from using the perceptual loss. First, the perceptual loss computes the distances between the estimated and ground-truth images in various feature levels, incorporating shallow, intermediate, and deep features and regulating the output of the network for high fidelity. Second, compared with the commonly-used mean-squared-error, the perceptual loss can effectively overcome the so-called over-smoothing problem~\cite{dosovitskiy2016generating}.

The objective function for the discriminator in Eq.~\eqref{equ_5} is composed of the Wasserstein distance (the first two terms) and the gradient penalty (the third term):
\begin{align}\label{equ_5}
    \mathcal{L}_{D} = \sum_{l=1}^{L} \Big[&\mathbb{E}(D(\widehat{\mat{I}}_{l})) - \mathbb{E}(D(\mat{I}_{l})) + \notag\\ &\lambda_{\rm gp}\mathbb{E}_{\tilde{\mat{I}}_{l}}\|\nabla_{\tilde{\mat{I}}_{l}}(D(\tilde{\mat{I}}_{l})) - 1 \|_2^2\Big],
\end{align}
where $\nabla$ is the gradient operator. $\tilde{\mat{I}}$ is a synthetic sample from real and fake samples and calculated as $\tilde{\mat{I}}=\epsilon\cdot\mat{I}+(1-\epsilon)\cdot G(\widehat{\mat{I}})$, where $\epsilon$ is a random value from the uniform distribution $\mathcal{U}:[0,1]$, and $\lambda_{\rm gp}$ is set to 10. Eq.~\eqref{equ_5} also accounts for all generator outputs and their corresponding ground truths.

\subsection{Frame interpolation}
Beside motion artifact reduction, our network can be used for frame interpolation; \emph{i.e.}, predicting missing frames given an incomplete motion blurred cardiac MRI sequence. In the network training and testing processes, frames were randomly excluded from the sequence so that an incomplete cine cardiac MRI sequence was created. Being different from the motion reduction task that differences between all outputs and their ground truths are taken into account, in our prediction experiment, only the discrepancy between the predicted frame and its ground truth is calculated.

\begin{figure}[t]
\centering
\includegraphics[width=3.5in]{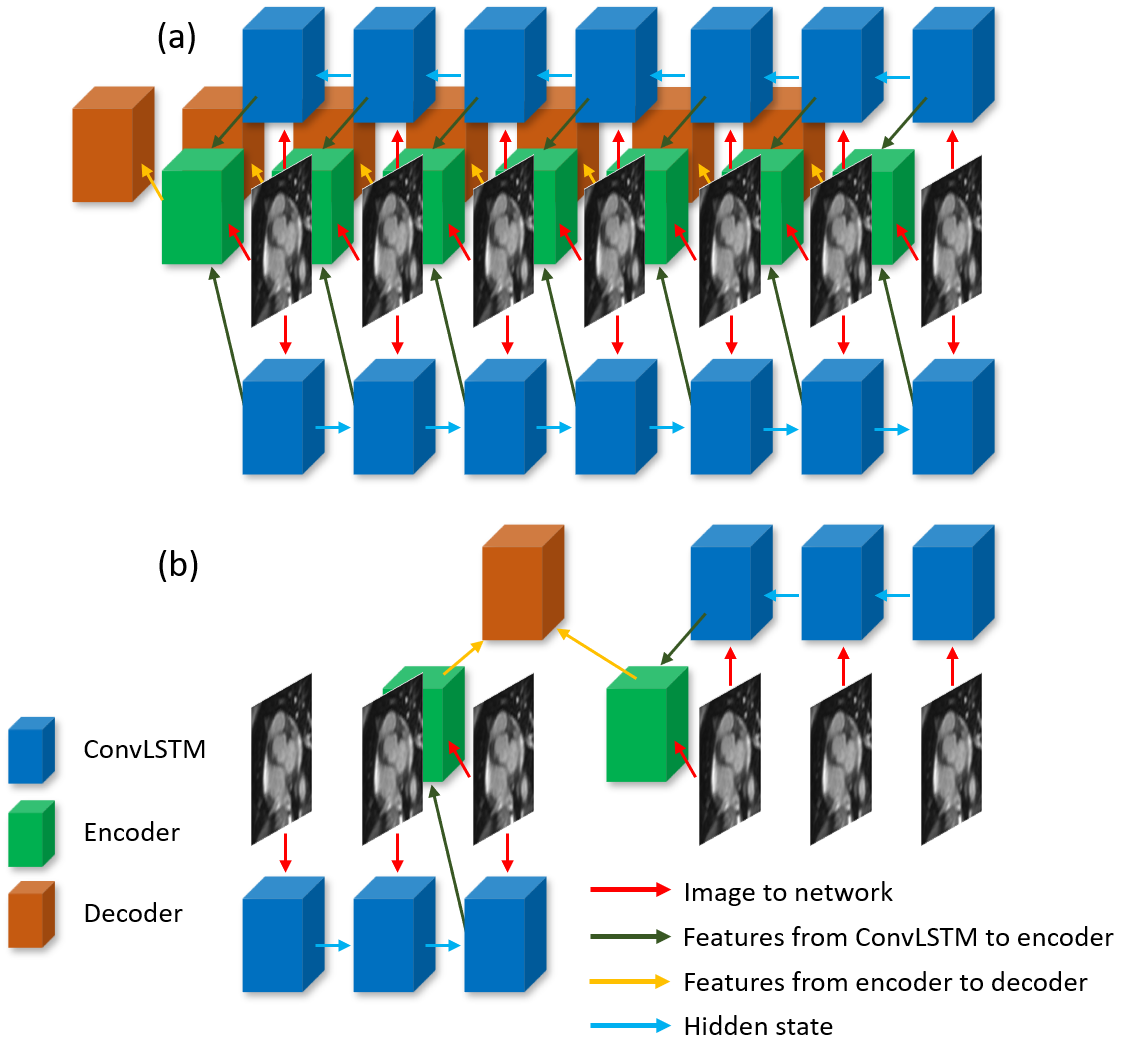}
\caption{Schematic views of two experimental designs. (a) Motion reduction and (b) frame prediction.}
\label{fig_3}
\end{figure}

\subsection{Implementation details}
In producing motion-blurred images, $N$ in Eq.~\eqref{equ_1} was set to 7 so that 15 adjacent frames were used to form each motion-blurred image. In the training process for motion artifact reduction, the length of an MRI sequence was set to 7. Each frame in the sequence is fed into the encoder, and the resultant features are concatenated with those generated from the ConvLSTM branches in each frame state. However, for the frame interpolation training, the length of a sequence was reduced to 6 with the central frame being excluded. Inputs to the encoder were the two adjacent frames of the missing frame of interest. The first three frames were input into the forward ConvLSTM branch and the last three frames into the backward branch. Only the features in the last state are passed into the encoder. The two experimental designs are illustrated in Fig.~\ref{fig_3}. In all experiments, the Adam optimizer was used with a learning rate of $10^{-4}$. We jointly trained the two ConvLSTM branches and the encoder-decoder network in an end-to-end manner. The training process continued for 50 epochs with a mini-batch size of 2.

We also try to implement our method on the Cedars dataset. Due to the small amount of scans, we haven't trained a new network on the Cedars datset. Instead, we choose the trained network on the ACDC dataset and fine-tuned the network using 10 randomly selected Cedars slices for 10 epoches. The learning rate was set to $2 \times 10^{-5}$. All experiments were conducted using PyTorch on a GTX 1080Ti GPU. We will make the source code public;y available after this paper's acceptance. 

\begin{figure}[!hbt]
\centering
\includegraphics[width=3.5in]{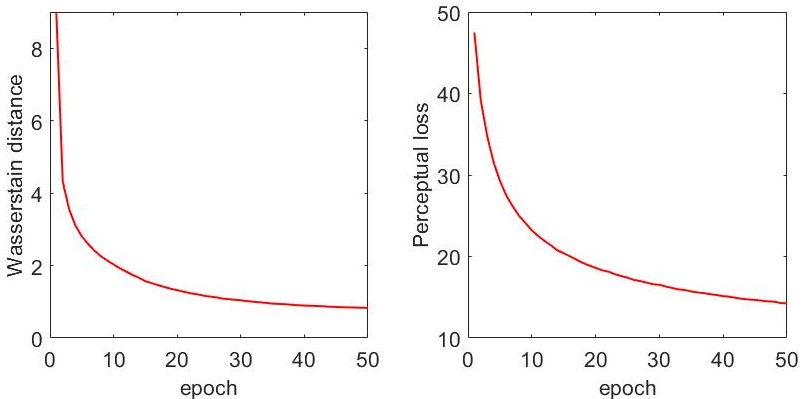}
\caption{Plots of the Wasserstein distance and the perceptual loss with respect to the number of epochs during the training process.}
\label{fig_4}
\end{figure}

\begin{figure}[!htb]
\centering
\includegraphics[width=3.5in]{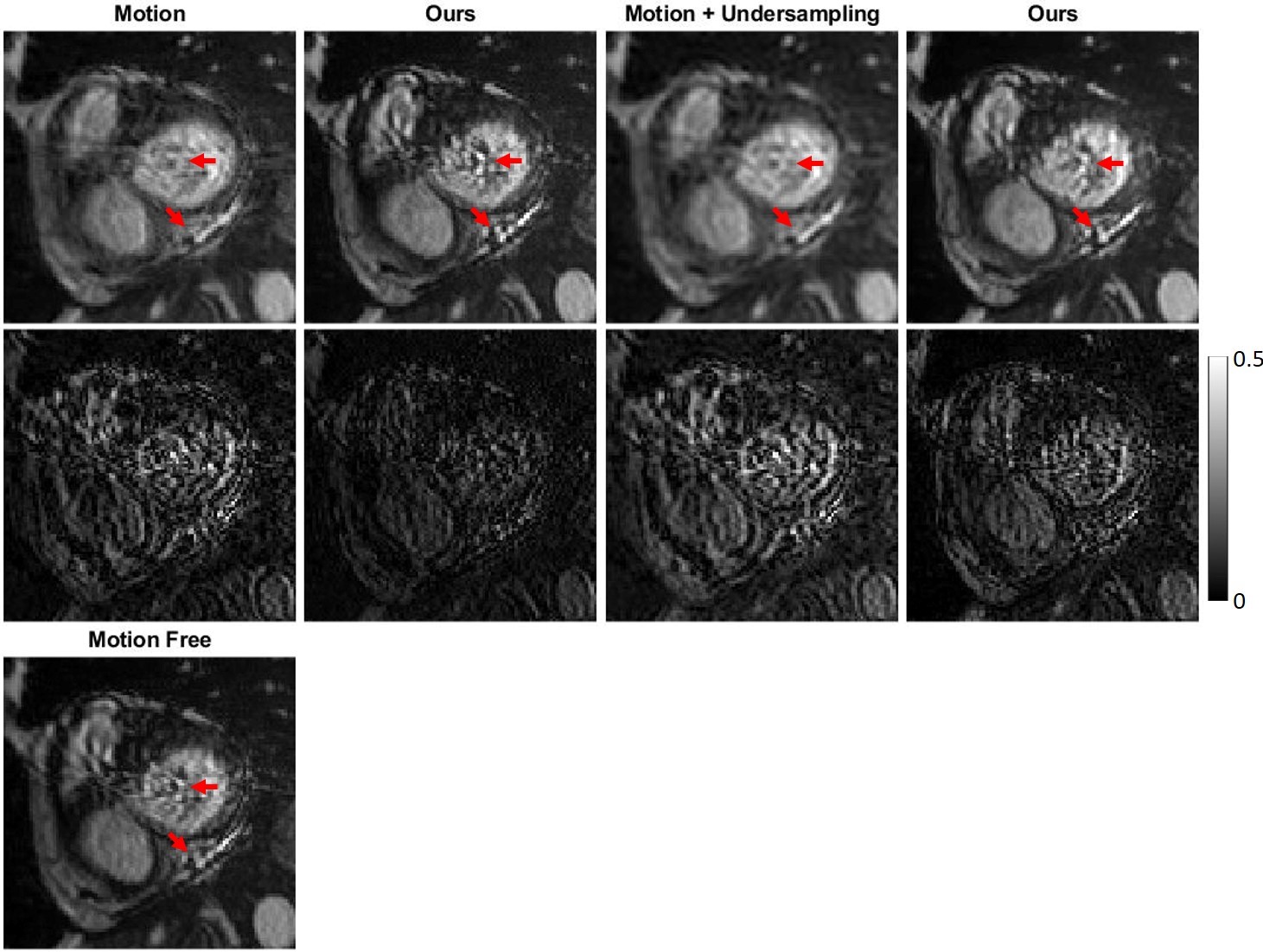}
\caption{Ablation study on datasets with only motion blurring versus both motion blurring and \textit{k}-space under-sampling.}
\label{fig_5}
\end{figure}

\begin{figure*}[!hbt]
\centering
\includegraphics[width=7in]{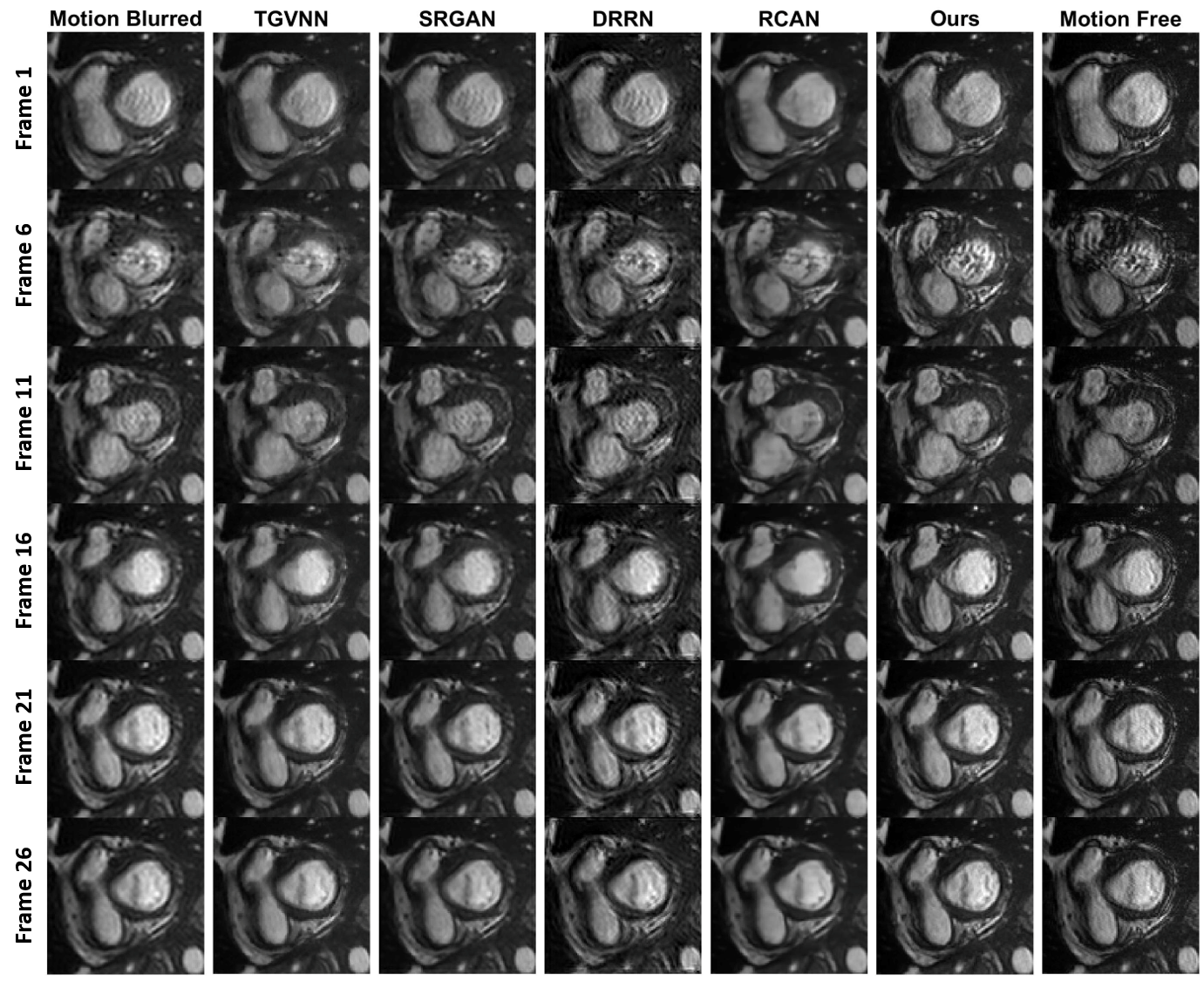}
\caption{Motion artifact reduction on the ACDC dataset, with representative six deblurred frames from the same image sequence.}
\label{fig_6}
\end{figure*}

\begin{figure*}[!hbt]
\centering
\includegraphics[width=7in]{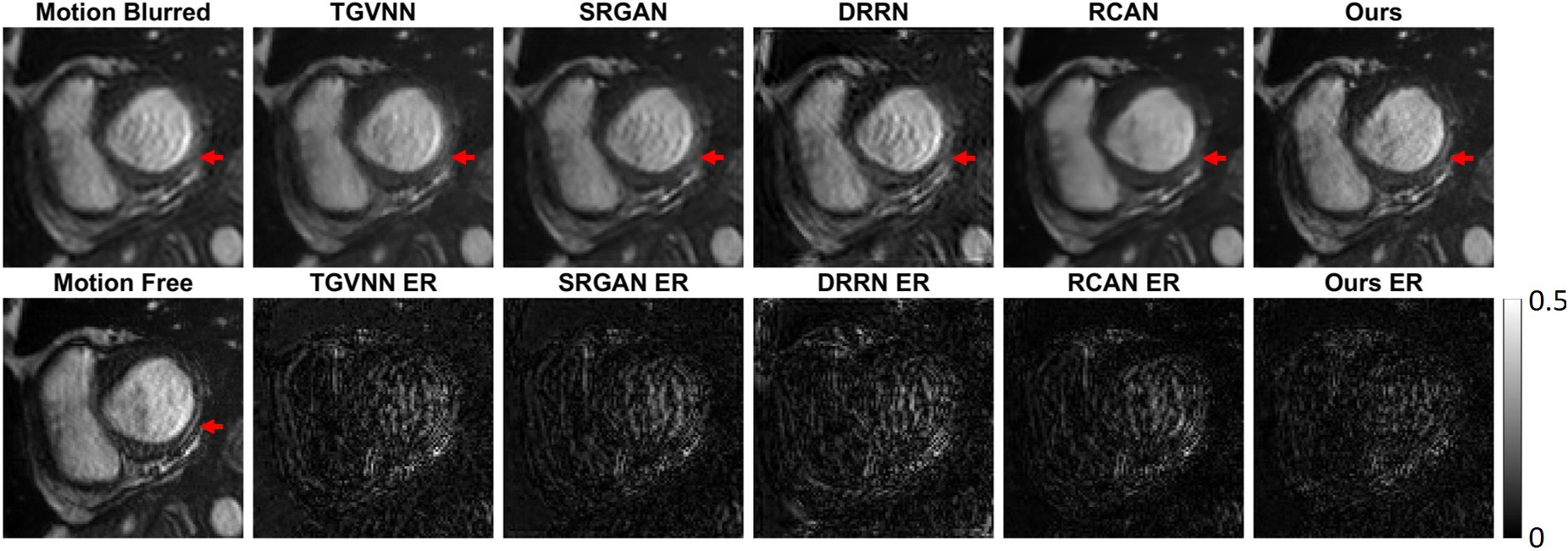}
\caption{Error analysis on the motion deblurred frame 1 in Fig.~\ref{fig_6}.}
\label{fig_7}
\end{figure*}

\section{Results}
\subsection{Convergence behavior}
Fig.~\ref{fig_4} shows the curves of the Wasserstein distance and the perceptual loss during the whole training process. These metrics demonstrate the convergence behavior of the training process of the proposed network. The Wasserstein distance shows the difference between the generator's outputs and the corresponding ground truth images. The perceptual loss represents the pixel-wise difference measured in the low, intermediate, and high level feature spaces. The lower the Wasserstein distance and perceptual loss values, the better learned MRI results. It can be observed that both of these two curves plummet in the initial training stage and then show gradually diminishing returns, indicating that our network was trained to a stable state to perform effectively.

\begin{figure*}[!htb]
\centering
\includegraphics[width=7in]{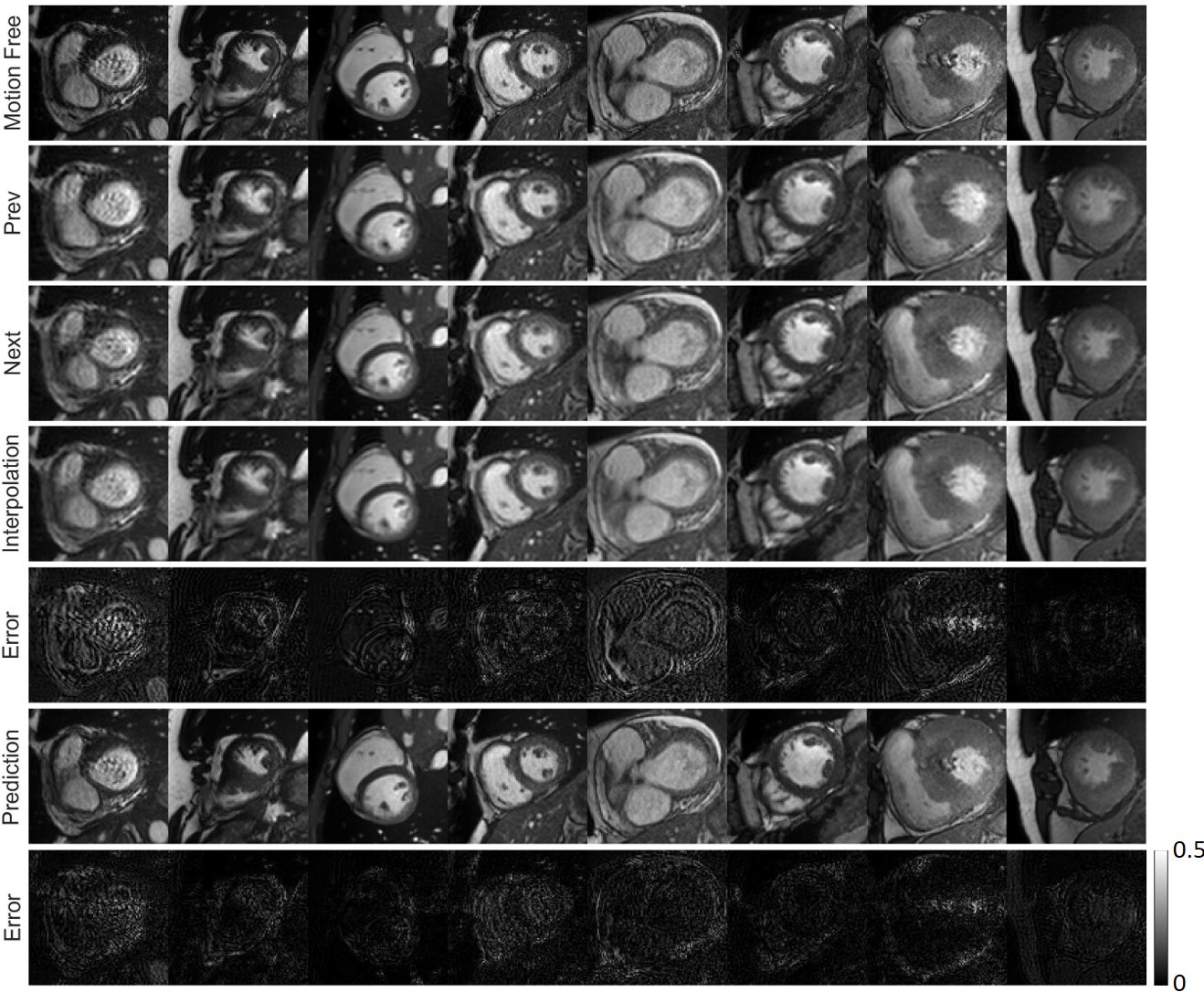}
\caption{Frame prediction results on the ACDC dataset. First row: motion-free ground truth images. Second row: previous motion-blurred frames. Third row: next motion-blurred frames. Fourth row: simple average interpolation frames. Fifth row: error maps between simple average interpolation frames and motion-free frames. Six row: our network-predicted frames. Seventh row: error maps between our predicted frames and motion-free frames. All the error maps are shown on the same grey scale bar.}
\label{fig_8}
\end{figure*}

\subsection{Ablation study}
We conducted an ablation study with two settings. The first ablation setting is to verify the effectiveness of our network structure. The second setting is to investigate our network's ability to reduce motion blurring.

In the first ablation setting, structural similarity (SSIM) and peak-to-noise ratio (PSNR) were used to measure the importance of three components in the proposed network: 1) encoder-decoder backbone, 2) two ConvLSTM branches, and 3) multi-scale analysis. Table~\ref{table_1} shows that the network with all these three components can generate results with the highest SSIM and PSNR scores, indicating the necessity of each component in the proposed network.

In the second ablation setting, we compared our results on two different datasets: 1) one with only motion blurring and 2) the other with both motion blurring and \textit{k}-space undersampling. For the dataset with only motion blurring, there was no zero-padding undersampling, and images were acquired directly from the 15-frame mixed \textit{k}-space data via the inverse Fourier transform. As shown in Fig.~\ref{fig_5}, on both datasets our network generated results with smaller errors when compared to the motion free ground truth. Notably, in the first two columns, it can be found that our network outputs can generate very high quality results close to the corresponding motion free ground truth. As only motion blurring exists in the network input, this finding strongly supports our hypothesis that the proposed network can reduce motion artifacts effectively. 

% ================== Table I =======================
\begin{table}[!h]
\caption{Averaged results from the ablation study on the ACDC dataset.}
\label{table_1}
\centering
\begin{adjustbox}{max width=1.0\columnwidth}
\begin{tabular}{ c c c c c}
\toprule
{Encoder-decoder} & {ConvLSTM} & {Multi-scale} & {SSIM} & {PSNR}\\  
\midrule
\checkmark & & & 0.816 & 26.294 \\
\checkmark & \checkmark & & 0.871 & 27.855 \\ 
\checkmark & & \checkmark & 0.821 & 26.487 \\
\checkmark & \checkmark & \checkmark & \textbf{0.884} & \textbf{28.514} \\
\bottomrule
\end{tabular}
\end{adjustbox}
\end{table}

\subsection{Motion artifact reduction}
We compared our motion artifact reduction results with four state-of-the-art methods: 1) total generalized variation and the nuclear norm (TGVNN)~\cite{wang2020dynamic}, 2) super-resolution using a generative adversarial network (SRGAN)~\cite{ledig2017photo}, 3) deep recursive residual network (DRRN)~\cite{tai2017image}, and 4) very deep residual channel attention network (RCAN)~\cite{zhang2018image}. TVGNN is a compressed sensing method for dynamic MRI acceleration, which uses the nuclear norm to encode the multi-frame time‐coherent information. SRGAN, DRRN, and RCAN are neural networks used to implement single image super-resolution imaging. According to the typical results in Figs.~\ref{fig_6},~\ref{fig_7}, and Table~\ref{table_2}, it can be observed that our proposed method achieves excellent results closest to the ground truth, with the highest SSIM and PSNR scores among all the mentioned methods. For example, Fig.~\ref{fig_7} shows that the edge-like features in the left ventricle wall can be well represented in our results but not with the other methods. Furthermore, our results have the least absolute errors relative to the ground truth among all the results.

% ================== Table II =======================
\begin{table}[!h]
\caption{Comparison between our motion reduction network and four existing methods on the ACDC dataset (MEAN$\pm$SD), where ``MEAN'' and ``SD'' represent the average value and the standard deviation, respectively.}
\label{table_2}
\centering
\begin{adjustbox}{max width=1.0\textwidth}
% \begin{tabular}{c c *{2}{S[separate-uncertainty=true, table-format=2.3(3), table-column-width=7.5em]}}
\begin{tabular}{c c c}
\toprule
{} & {SSIM} & {PSNR} \\
\midrule
{TGVNN} & 0.786 $\pm$ 0.051 & 26.475 $\pm$ 2.063 \\
{SRGAN} & 0.832 $\pm$ 0.063 & 27.136 $\pm$ 3.535 \\
{DRRN}  & 0.817 $\pm$ 0.052 & 26.866 $\pm$ 1.555 \\
{RCAN}  & 0.841 $\pm$ 0.051 & 27.210 $\pm$ 2.287 \\
{Ours}  & \textbf{0.884 $\pm$ 0.047} & \textbf{28.514 $\pm$ 2.210} \\
\bottomrule
\end{tabular}
\end{adjustbox}
\end{table}

\subsection{Missing frame estimation}
Fig.~\ref{fig_8} presents the frame interpolation results. Given an incomplete motion-blurred cine cardiac MRI sequence, our network can predict missing frames, clearly outperforming the linear interpolation of the previous and next frames. It can be seen that the position and shape of the ventricle structures in our predicted missing frames are quite close to the target ground truth frame. In addition, compared to the previous and next frames, the image quality of the prediction is greatly improved.

\begin{figure}[!htb]
\centering
\includegraphics[width=3.5in]{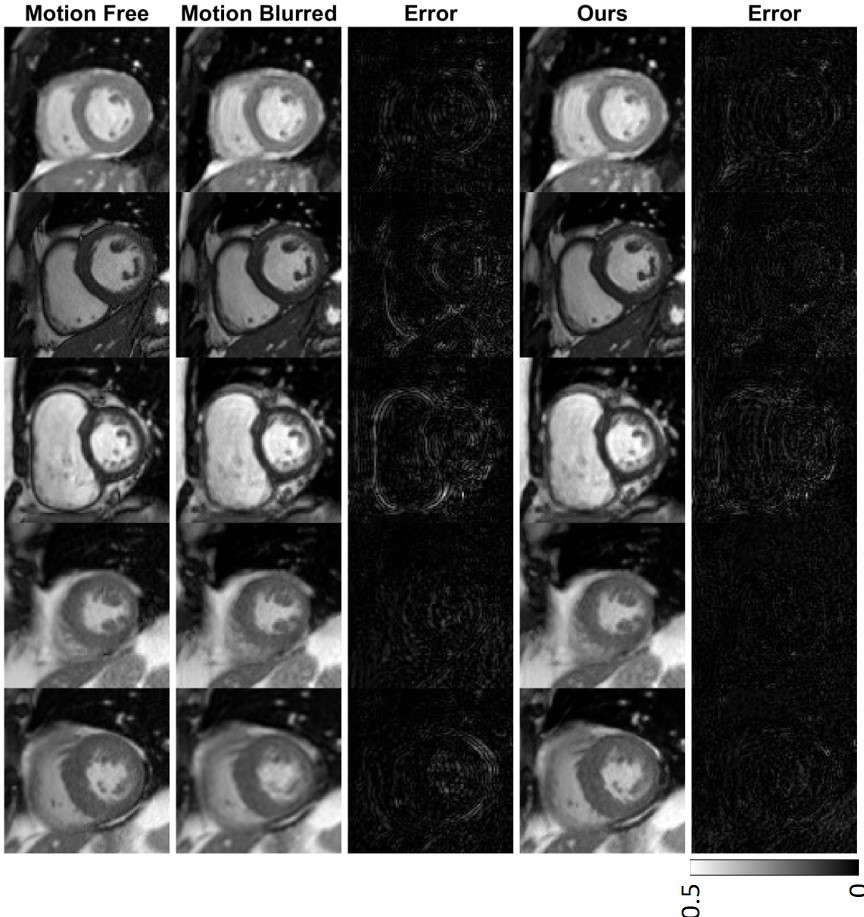}
\caption{Motion artifacts reduction results on the Cedars dataset. First column: motion-free ground truths. Second column: motion-blurred frames based on frame mixture and zero-padding. Third column: error maps between motion-blurred frames and motion-free counterparts. Fourth column: our results. Fifth column: error maps between our results and the ground truth. All error maps are shown with the same grey scale bar.}
\label{fig_9}
\end{figure}

\subsection{Generalization to the Cedars dataset}
To investigate our network's feasibility on more generalized scanners, we further implement our method on the Cedars dataset. Both the Cedars and ACDC datasets contain cine cardiac MRI sequences with a maximum length of 30. However, the main difference between these datasets lies in the MRI scanners used to acquire them, resulting in different spatial resolution and slice thickness. It is clear in Fig.~\ref{fig_9} that our results show fewer blurred edges than the original motion blurred images. SSIM (0.870$\pm$0.031) and PSNR (28.809$\pm$2.919) scores of our results are significantly better than that of the motion blurred images (SSIM: 0.855$\pm$0.047, PSNR: 26.423$\pm$3.958).

\begin{figure}[!htb]
\centering
\includegraphics[width=3.5in]{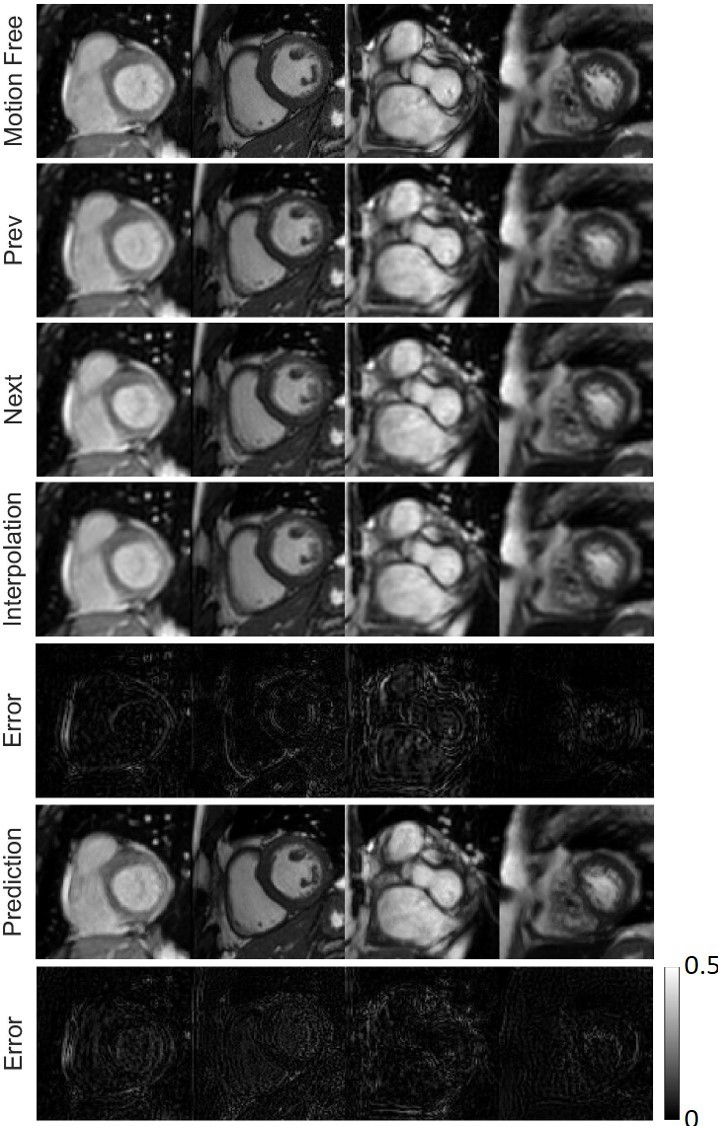}
\caption{Missing frame prediction results on the Cedars dataset. First row: motion-free ground truths. Second row: previous motion-blurred frames. Third row: next motion-blurred frames. Fourth row: linearly interpolated frames. Fifth row: error maps between linear interpolation and ground truth. Six row: our network-predicted frames. Seventh row: error maps between our prediction and ground truth.}
\label{fig_10}
\end{figure}

Fig.~\ref{fig_10} shows missing frame estimation results based on the Cedars dataset. Similar to results on the ACDC dataset, the prediction results in Fig.~\ref{fig_10} show much clearer tissue shapes and quite accurate tissue structures than the blurred images, which turns out that our method can improve the temporal resolution.

\subsection{Motion deblurring from 5-frame mixture}
\begin{figure}[!htb]
\centering
\includegraphics[width=3.5in]{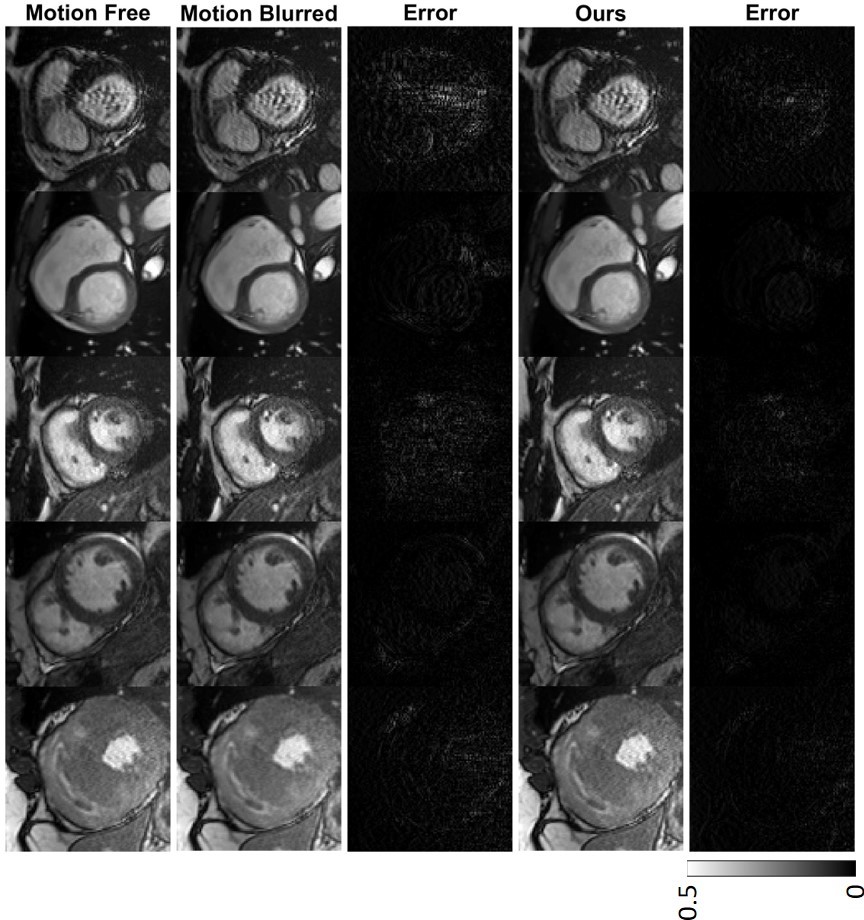}
\caption{Results on another clinical dataset. First column: motion-free ground truths. Second column: motion-blurred frames via 5-frame mixture. Third column: error maps between motion-blurred frames and the ground truth. Fourth column: our results. Fifth column: error maps between our results and the ground truth. All error maps are shown on the same grey scale bar.}
\label{fig_11}
\end{figure}

Our results in the previous subsections are based on motion blurred datasets realistically created by fusing \textit{k}-space data from 15 adjacent frames. These datasets are with evident motion blurring and are technically challenging for motion artifact reduction. Our previous results show the effectiveness of our proposed network in solving such a challenging task. However, clinically acquired cine cardiac MRI sequences can be with less motion blurring than the datasets used in this paper. For a typical cine cardiac MRI scan, the temporal resolution is between the range of around 50 and 100 ms~\cite{wintersperger2003single}. With such temporal resolution, the motion blurred images should only contain \textit{k}-space data within the adjacent 3\text{$\sim$}5 frames. To apply our method to a less challenging clinical dataset, we created a dataset in which each frame was made by fusing \textit{k}-space data from 5 adjacent frames. Results are shown in Fig.~\ref{fig_11}. Again, our results are much closer to the corresponding ground truth images than the original motion-blurred frames.

\section{Discussions}
In this paper, we proposed an advanced recurrent neural network for  cardiac cine MRI motion reduction. According to Table~\ref{table_1}, the results with both ConvLSTM branches and multi-scale structures have the highest SSIM and PSNR scores. Compared with competing networks that only extract spatial features (like the encoder-decoder, SRGAN, DRRN, and RCAN), our network fully utilizes forward and backward ConvLSTM mechanisms to simultaneously extract spatial and temporal features from an image sequence. Thus, more information can be extracted and utilized by our network for better reconstruction results than the competing methods. In particular, we performed multi-scale analysis to further improve the performance of our network. When compared with the encoder only using 3 $\times$ 3 convolution kernels to extract local features, multi-scale analysis boosted the feature extraction capability of the encoder so that both local and global features can be captured via convolution kernels with different size of perception fields. In addition, since we used ConvLSTM to extract temporal features, it is feasible to make inference with any sequence length, which improves the flexibility of our method.

Image sequences in the ACDC dataset were acquired on two MRI scanners with different magnetic fields and hence different spatial resolution and slice thickness. We trained the network with a mixture of those image sequences in hope of improving the robustness of the proposed network so that it can be applied to sequences from multiple MRI scanners. Our testing results on the ACDC dataset suggested that our trained network is robust and enables satisfactory motion artifacts reduction and missing frame prediction on image sequences from two different MRI scanners. Moreover, Figs.~\ref{fig_9} and \ref{fig_10} also show satisfactory results in motion artifact reduction and missing frame prediction on the Cedars dataset, which further indicates the accuracy and robustness of our method. 

Our proposed network has potential clinical utilities. First, it can alleviate motion artifacts and recover the actual shape of cardiac structures like ventricles, which will be helpful to boost the diagnostic accuracy. Second, future cine cardiac MRI scans without ECG signals may be possible to obtain high quality motion-free cines as our method has established the feasibility of removing motion induced artifacts during the scan, leading to a simplified workflow of clinical cine cardiac MRI scanning with little pre-scan preparation. Third, our method can accelerate cine cardiac MRI scanning. According to our results, image quality can be greatly improved based on \textit{k}-space under-sampled MR images, future MRI scans could be accelerated four times or even more in this way. Furthermore, our results on predicting missing frames indicate that cine imaging can be enhanced for higher temporal resolution using our network. The proposed network can extract critical information from adjacent frames and interpolate missing frames intelligently. As a consequence, more frames than what are clinical obtained can be computationally acquired.

\section{Conclusion}
We have proposed an advanced recurrent neural network for cine cardiac MRI motion artifact reduction and acceleration as well as missing frame interpolation. We have utilized forward and backward ConvLSTM branches so that both spatial and temporal features can be extracted for synergistic imaging. To achieve better results, we have incorporated multi-scale analysis in the encoder. Compared with existing state-of-the-art methods, our method has achieved the highest SSIM and PSNR scores for motion artifact reduction. Apart from reducing motion artifacts, our method has also produced encouraging results in improving temporal resolution of cardiac cine by high-fidelity interpolation. Our results on the ACDC and Cedars datasets have consistently demonstrated the accuracy and robustness of our proposed network. Our method may help improve the image quality of future cine cardiac MRI scans and shorten the scan process significantly without compromising the image quality.

\ifCLASSOPTIONcaptionsoff
  \newpage
\fi

\bibliographystyle{ieeetr}
\bibliography{bibliography}

\begin{thebibliography}{10}

\bibitem{motwani2013mr}
M.~Motwani, A.~Kidambi, B.~A. Herzog, A.~Uddin, J.~P. Greenwood, and S.~Plein,
  ``Mr imaging of cardiac tumors and masses: a review of methods and clinical
  applications,'' {\em Radiology}, vol.~268, no.~1, pp.~26--43, 2013.

\bibitem{sechtem1987cine}
U.~Sechtem, P.~W. Pflugfelder, R.~D. White, R.~G. Gould, W.~Holt, M.~J. Lipton,
  and C.~B. Higgins, ``Cine mr imaging: potential for the evaluation of
  cardiovascular function,'' {\em American Journal of Roentgenology}, vol.~148,
  no.~2, pp.~239--246, 1987.

\bibitem{leiner2019machine}
T.~Leiner, D.~Rueckert, A.~Suinesiaputra, B.~Bae{\ss}ler, R.~Nezafat,
  I.~I{\v{s}}gum, and A.~A. Young, ``Machine learning in cardiovascular
  magnetic resonance: basic concepts and applications,'' {\em Journal of
  Cardiovascular Magnetic Resonance}, vol.~21, no.~1, p.~61, 2019.

\bibitem{larson2004self}
A.~C. Larson, R.~D. White, G.~Laub, E.~R. McVeigh, D.~Li, and O.~P. Simonetti,
  ``Self-gated cardiac cine mri,'' {\em Magnetic Resonance in Medicine: An
  Official Journal of the International Society for Magnetic Resonance in
  Medicine}, vol.~51, no.~1, pp.~93--102, 2004.

\bibitem{crowe2004automated}
M.~E. Crowe, A.~C. Larson, Q.~Zhang, J.~Carr, R.~D. White, D.~Li, and O.~P.
  Simonetti, ``Automated rectilinear self-gated cardiac cine imaging,'' {\em
  Magnetic Resonance in Medicine: An Official Journal of the International
  Society for Magnetic Resonance in Medicine}, vol.~52, no.~4, pp.~782--788,
  2004.

\bibitem{schmitt2008128}
M.~Schmitt, A.~Potthast, D.~E. Sosnovik, J.~R. Polimeni, G.~C. Wiggins,
  C.~Triantafyllou, and L.~L. Wald, ``A 128-channel receive-only cardiac coil
  for highly accelerated cardiac mri at 3 tesla,'' {\em Magnetic Resonance in
  Medicine: An Official Journal of the International Society for Magnetic
  Resonance in Medicine}, vol.~59, no.~6, pp.~1431--1439, 2008.

\bibitem{tsao2003k}
J.~Tsao, P.~Boesiger, and K.~P. Pruessmann, ``k-t blast and k-t sense: dynamic
  mri with high frame rate exploiting spatiotemporal correlations,'' {\em
  Magnetic Resonance in Medicine: An Official Journal of the International
  Society for Magnetic Resonance in Medicine}, vol.~50, no.~5, pp.~1031--1042,
  2003.

\bibitem{lingala2011accelerated}
S.~G. Lingala, Y.~Hu, E.~DiBella, and M.~Jacob, ``Accelerated dynamic mri
  exploiting sparsity and low-rank structure: kt slr,'' {\em IEEE transactions
  on medical imaging}, vol.~30, no.~5, pp.~1042--1054, 2011.

\bibitem{tremoulheac2014dynamic}
B.~Tr{\'e}moulh{\'e}ac, N.~Dikaios, D.~Atkinson, and S.~R. Arridge, ``Dynamic
  {MR} image reconstruction--separation from undersampled (\textbf{k},
  \textit{t})-space via low-rank plus sparse prior,'' {\em IEEE transactions on
  medical imaging}, vol.~33, no.~8, pp.~1689--1701, 2014.

\bibitem{otazo2015low}
R.~Otazo, E.~Candes, and D.~K. Sodickson, ``Low-rank plus sparse matrix
  decomposition for accelerated dynamic mri with separation of background and
  dynamic components,'' {\em Magnetic Resonance in Medicine}, vol.~73, no.~3,
  pp.~1125--1136, 2015.

\bibitem{christodoulou2018magnetic}
A.~G. Christodoulou, J.~L. Shaw, C.~Nguyen, Q.~Yang, Y.~Xie, N.~Wang, and
  D.~Li, ``Magnetic resonance multitasking for motion-resolved quantitative
  cardiovascular imaging,'' {\em Nature biomedical engineering}, vol.~2, no.~4,
  p.~215, 2018.

\bibitem{schloegl2017infimal}
M.~Schloegl, M.~Holler, A.~Schwarzl, K.~Bredies, and R.~Stollberger, ``Infimal
  convolution of total generalized variation functionals for dynamic {MRI},''
  {\em Magnetic resonance in medicine}, vol.~78, no.~1, pp.~142--155, 2017.

\bibitem{wang2020dynamic}
D.~Wang, D.~S. Smith, and X.~Yang, ``Dynamic {MR} image reconstruction based on
  total generalized variation and low-rank decomposition,'' {\em Magnetic
  Resonance in Medicine}, vol.~83, no.~6, pp.~2064--2076, 2020.

\bibitem{poudel2016recurrent}
R.~P. Poudel, P.~Lamata, and G.~Montana, ``Recurrent fully convolutional neural
  networks for multi-slice mri cardiac segmentation,'' in {\em Reconstruction,
  segmentation, and analysis of medical images}, pp.~83--94, Springer, 2016.

\bibitem{tan2017convolutional}
L.~K. Tan, Y.~M. Liew, E.~Lim, and R.~A. McLaughlin, ``Convolutional neural
  network regression for short-axis left ventricle segmentation in cardiac cine
  mr sequences,'' {\em Medical image analysis}, vol.~39, pp.~78--86, 2017.

\bibitem{romaguera2017left}
L.~V. Romaguera, M.~G.~F. Costa, F.~P. Romero, and C.~F.~F. Costa~Filho, ``Left
  ventricle segmentation in cardiac mri images using fully convolutional neural
  networks,'' in {\em Medical Imaging 2017: Computer-Aided Diagnosis},
  vol.~10134, p.~101342Z, International Society for Optics and Photonics, 2017.

\bibitem{luo2016deep}
G.~Luo, R.~An, K.~Wang, S.~Dong, and H.~Zhang, ``A deep learning network for
  right ventricle segmentation in short-axis mri,'' in {\em 2016 Computing in
  Cardiology Conference (CinC)}, pp.~485--488, IEEE, 2016.

\bibitem{de2017end}
B.~D. de~Vos, F.~F. Berendsen, M.~A. Viergever, M.~Staring, and I.~I{\v{s}}gum,
  ``End-to-end unsupervised deformable image registration with a convolutional
  neural network,'' in {\em Deep Learning in Medical Image Analysis and
  Multimodal Learning for Clinical Decision Support}, pp.~204--212, Springer,
  2017.

\bibitem{shan2020synergizing}
H.~Shan, X.~Jia, P.~Yan, Y.~Li, H.~Paganetti, and G.~Wang, ``Synergizing
  medical imaging and radiotherapy with deep learning,'' {\em Machine Learning:
  Science and Technology}, 2020.

\bibitem{lyu2019super}
Q.~Lyu, C.~You, H.~Shan, Y.~Zhang, and G.~Wang, ``Super-resolution {MRI} and
  {CT} through {GAN-CIRCLE},'' in {\em Developments in X-Ray Tomography XII},
  vol.~11113, p.~111130X, International Society for Optics and Photonics, 2019.

\bibitem{chaudhari2018super}
A.~S. Chaudhari, Z.~Fang, F.~Kogan, J.~Wood, K.~J. Stevens, E.~K. Gibbons,
  J.~H. Lee, G.~E. Gold, and B.~A. Hargreaves, ``Super-resolution
  musculoskeletal {MRI} using deep learning,'' {\em Magnetic resonance in
  medicine}, vol.~80, no.~5, pp.~2139--2154, 2018.

\bibitem{chen2018efficient}
Y.~Chen, F.~Shi, A.~G. Christodoulou, Y.~Xie, Z.~Zhou, and D.~Li, ``Efficient
  and accurate {MRI} super-resolution using a generative adversarial network
  and 3{D} multi-level densely connected network,'' in {\em International
  Conference on Medical Image Computing and Computer-Assisted Intervention},
  pp.~91--99, Springer, 2018.

\bibitem{lyu2020multi}
Q.~Lyu, H.~Shan, C.~Steber, C.~Helis, C.~T. Whitlow, M.~Chan, and G.~Wang,
  ``Multi-contrast super-resolution {MRI} through a progressive network,'' {\em
  IEEE Transactions on Medical Imaging}, 2020.

\bibitem{lyu2020mri}
Q.~Lyu, H.~Shan, and G.~Wang, ``{MRI} super-resolution with ensemble learning
  and complementary priors,'' {\em IEEE Transactions on Computational Imaging},
  vol.~6, pp.~615--624, 2020.

\bibitem{liu2018fusing}
C.~Liu, X.~Wu, X.~Yu, Y.~Tang, J.~Zhang, and J.~Zhou, ``Fusing multi-scale
  information in convolution network for {MR} image super-resolution
  reconstruction,'' {\em Biomedical engineering online}, vol.~17, no.~1,
  p.~114, 2018.

\bibitem{zeng2018simultaneous}
K.~Zeng, H.~Zheng, C.~Cai, Y.~Yang, K.~Zhang, and Z.~Chen, ``Simultaneous
  single-and multi-contrast super-resolution for brain {MRI} images based on a
  convolutional neural network,'' {\em Computers in biology and medicine},
  vol.~99, pp.~133--141, 2018.

\bibitem{yang2018multi}
R.~Yang, M.~Xu, Z.~Wang, and T.~Li, ``Multi-frame quality enhancement for
  compressed video,'' in {\em Proceedings of the IEEE Conference on Computer
  Vision and Pattern Recognition}, pp.~6664--6673, 2018.

\bibitem{guan2019mfqe}
Z.~Guan, Q.~Xing, M.~Xu, R.~Yang, T.~Liu, and Z.~Wang, ``{MFQE} 2.0: A new
  approach for multi-frame quality enhancement on compressed video,'' {\em
  arXiv preprint arXiv:1902.09707}, 2019.

\bibitem{xue2019video}
T.~Xue, B.~Chen, J.~Wu, D.~Wei, and W.~T. Freeman, ``Video enhancement with
  task-oriented flow,'' {\em International Journal of Computer Vision},
  vol.~127, no.~8, pp.~1106--1125, 2019.

\bibitem{schlemper2017deep}
J.~Schlemper, J.~Caballero, J.~V. Hajnal, A.~N. Price, and D.~Rueckert, ``A
  deep cascade of convolutional neural networks for dynamic {MR} image
  reconstruction,'' {\em IEEE transactions on Medical Imaging}, vol.~37, no.~2,
  pp.~491--503, 2017.

\bibitem{qin2018convolutional}
C.~Qin, J.~Schlemper, J.~Caballero, A.~N. Price, J.~V. Hajnal, and D.~Rueckert,
  ``Convolutional recurrent neural networks for dynamic mr image
  reconstruction,'' {\em IEEE transactions on medical imaging}, vol.~38, no.~1,
  pp.~280--290, 2018.

\bibitem{luo2016novel}
G.~Luo, G.~Sun, K.~Wang, S.~Dong, and H.~Zhang, ``A novel left ventricular
  volumes prediction method based on deep learning network in cardiac {MRI},''
  in {\em 2016 Computing in Cardiology Conference (CinC)}, pp.~89--92, IEEE,
  2016.

\bibitem{zhang2019spatio}
L.~Zhang, L.~Lu, X.~Wang, R.~M. Zhu, M.~Bagheri, R.~M. Summers, and J.~Yao,
  ``Spatio-temporal convolutional {LSTM}s for tumor growth prediction by
  learning 4{D} longitudinal patient data,'' {\em IEEE Transactions on Medical
  Imaging}, 2019.

\bibitem{bello2019deep}
G.~A. Bello, T.~J. Dawes, J.~Duan, C.~Biffi, A.~De~Marvao, L.~S. Howard,
  J.~S.~R. Gibbs, M.~R. Wilkins, S.~A. Cook, D.~Rueckert, {\em et~al.},
  ``Deep-learning cardiac motion analysis for human survival prediction,'' {\em
  Nature machine intelligence}, vol.~1, no.~2, pp.~95--104, 2019.

\bibitem{bernard2018deep}
O.~Bernard, A.~Lalande, C.~Zotti, F.~Cervenansky, X.~Yang, P.-A. Heng,
  I.~Cetin, K.~Lekadir, O.~Camara, M.~A.~G. Ballester, {\em et~al.}, ``Deep
  learning techniques for automatic {MRI} cardiac multi-structures segmentation
  and diagnosis: is the problem solved,'' {\em IEEE transactions on medical
  imaging}, vol.~37, no.~11, pp.~2514--2525, 2018.

\bibitem{xingjian2015convolutional}
S.~Xingjian, Z.~Chen, H.~Wang, D.-Y. Yeung, W.-K. Wong, and W.-c. Woo,
  ``Convolutional {LSTM} network: A machine learning approach for precipitation
  nowcasting,'' in {\em Advances in neural information processing systems},
  pp.~802--810, 2015.

\bibitem{johnson2016perceptual}
J.~Johnson, A.~Alahi, and L.~Fei-Fei, ``Perceptual losses for real-time style
  transfer and super-resolution,'' in {\em European conference on computer
  vision}, pp.~694--711, Springer, 2016.

\bibitem{simonyan2014very}
K.~Simonyan and A.~Zisserman, ``Very deep convolutional networks for
  large-scale image recognition,'' {\em arXiv preprint arXiv:1409.1556}, 2014.

\bibitem{dosovitskiy2016generating}
A.~Dosovitskiy and T.~Brox, ``Generating images with perceptual similarity
  metrics based on deep networks,'' in {\em Advances in neural information
  processing systems}, pp.~658--666, 2016.

\bibitem{ledig2017photo}
C.~Ledig, L.~Theis, F.~Husz{\'a}r, J.~Caballero, A.~Cunningham, A.~Acosta,
  A.~Aitken, A.~Tejani, J.~Totz, Z.~Wang, {\em et~al.}, ``Photo-realistic
  single image super-resolution using a generative adversarial network,'' in
  {\em Proceedings of the IEEE conference on computer vision and pattern
  recognition}, pp.~4681--4690, 2017.

\bibitem{tai2017image}
Y.~Tai, J.~Yang, and X.~Liu, ``Image super-resolution via deep recursive
  residual network,'' in {\em Proceedings of the IEEE conference on computer
  vision and pattern recognition}, pp.~3147--3155, 2017.

\bibitem{zhang2018image}
Y.~Zhang, K.~Li, K.~Li, L.~Wang, B.~Zhong, and Y.~Fu, ``Image super-resolution
  using very deep residual channel attention networks,'' in {\em Proceedings of
  the European Conference on Computer Vision (ECCV)}, pp.~286--301, 2018.

\bibitem{wintersperger2003single}
B.~J. Wintersperger, K.~Nikolaou, O.~Dietrich, J.~Rieber, M.~Nittka, M.~F.
  Reiser, and S.~O. Schoenberg, ``Single breath-hold real-time cine mr imaging:
  improved temporal resolution using generalized autocalibrating partially
  parallel acquisition (grappa) algorithm,'' {\em European radiology}, vol.~13,
  no.~8, pp.~1931--1936, 2003.

\end{thebibliography}

\end{document}